\documentstyle[multicol,aps,prl,psfig,graphicx,array]{revtex}

\renewcommand{\narrowtext}{\begin{multicols}{2} \global\columnwidth20.5pc}
\def\al{\alpha}
\def\be{\beta}
\def\ga{\gamma}
\def\de{\delta}
\def\ep{\epsilon}

\def\et{\eta}

\def\ka{\kappa}
\def\la{\lambda}

\def\rh{\rho}

\def\si{\sigma}

\def\ps{\psi}
\def\om{\omega}
\def\Ga{\Gamma}

\def\Si{\Sigma}

\def\fr#1#2{{{#1} \over {#2}}}
\def\frac#1#2{\textstyle{{{#1} \over {#2}}}}
\def\half{{\textstyle{1\over 2}}}
\def\pt#1{\phantom{#1}}
\def\prt{\partial}

\def\lsim{\mathrel{\rlap{\lower4pt\hbox{\hskip1pt$\sim$}}
    \raise1pt\hbox{$<$}}}
\def\gsim{\mathrel{\rlap{\lower4pt\hbox{\hskip1pt$\sim$}}
    \raise1pt\hbox{$>$}}}

\newcommand{\bequ}{\begin{equation}}
\newcommand{\eequ}{\end{equation}}
\newcommand{\beq}{\begin{eqnarray}}
\newcommand{\eeq}{\end{eqnarray}}
\newcommand{\bea}{\begin{eqnarray}}
\newcommand{\eea}{\end{eqnarray}}
\newcommand{\rf}[1]{(\ref{#1})}

\def\kaf{{(k_{AF})}}
\def\kf{{(k_{F})}}
\def\kkaf{{$(k_{AF})_\mu$}}
\def\kkf{{$(k_{F})_{\ka\la\mu\nu}$}}

\def\a{$a_\mu$}
\def\b{$b_\mu$}
\def\c{$c_{\mu\nu}$}
\def\d{$d_{\mu\nu}$}
\def\e{$e_\mu$}
\def\f{$f_\mu$}
\def\g{$g_{\la\mu\nu}$}
\def\H{$H_{\mu\nu}$}

\newcommand{\eqref}[1]{(\ref{#1})}

\newcommand{\fslash}[1]{#1 \!\!\! /}

\newcommand{\incps}[5]{\includegraphics[#2,#3][#4,#5]{#1}}

\newcommand{\incpicwh}[3]{\includegraphics[width=#2,height=#3]{#1}}

\newcommand{\pad}{\hspace{-0.5cm}}
\newcommand{\remark}[1]{}
\newcommand{\fgl}[1]{\hspace{1.05cm}#1\hspace{-1.05cm}}
\newcommand{\qedpic}[1]{\incpicwh{#1}{2.6cm}{2.6cm}}
\newcommand{\negpad}{\!\!\!\!\!\!\!\!}

\begin{document}

\title{One-Loop Renormalization of Lorentz-Violating Electrodynamics}
\author{V.\ Alan Kosteleck\'y,$^a$
Charles D.\ Lane,$^b$ 
and Austin G.M.\ Pickering$^a$}
\address{$^a$Physics Department, Indiana University, 
          Bloomington, IN 47405, U.S.A.}
\address{$^b$Physics Department, Berry College, 
          Mount Berry, GA 30149, U.S.A.}
\date{IUHET 442, October 2001; 
accepted for publication in Physical Review D} 
\maketitle

\begin{abstract}
We show that the general Lorentz- and CPT-violating extension 
of quantum electrodynamics is one-loop renormalizable.
The one-loop Lorentz-violating beta functions are obtained, 
and the running of the coefficients for Lorentz and CPT violation 
is determined.
Some implications for theory and experiment are discussed.
\end{abstract}


\narrowtext

\section{Introduction}

The standard model of particle physics is invariant
under Lorentz and CPT transformations.
However,
the possibility that nature exhibits
small violations of Lorentz and CPT symmetry 
appears compatible with quantum field theory
and with existing experiments
\cite{cpt98}.
A general description of the associated effects
can be formulated at the level of quantum field theory
as a Lorentz- and CPT-violating standard-model extension 
\cite{ck}.
The lagrangian of this theory includes all possible operators 
that are observer Lorentz scalars
and that are formed from standard-model fields
and coupling coefficients with Lorentz indices.
Imposing the usual SU(3)$\times$SU(2)$\times$U(1) gauge invariance
and restricting attention to low-energy effects,
the standard-model extension
is well approximated by the usual standard model
together with all possible Lorentz-violating terms 
of mass dimension four or less
that are constructed from standard-model fields.

Among the interesting open issues associated with
Lorentz and CPT violation 
is the manner in which the low-energy theory
connects to the underlying Planck-level theory  
as the energy scale is increased. 
Some insight into this link has been obtained
through the study of causality and stability
in Lorentz-violating quantum field theory
\cite{kle}.
In the present work, we study a different facet of this connection,
involving the role of radiative corrections and the renormalization group.

To provide a definite focus and a tractable scope,
we limit attention here to the special case
of effects from one-loop divergences
in the Lorentz- and CPT-violating quantum electrodynamics (QED) 
of a single fermion.
This QED extension can be regarded as a specific limit 
of the standard-model extension.
Even in this simplified limit,
relatively little is known about loop effects.
Some one-loop calculations have been performed
in the photon sector
\cite{ck,jk},
but a comprehensive treatment has been lacking.
One goal of the present work is to fill this gap.
Tools such as a generalization of the Furry theorem 
\cite{wf}
are developed,
and all divergent one-loop corrections are determined.
We use these to prove one-loop renormalizability
and gauge invariance of the theory.
The calculations are presented here in dimensional regularization
\cite{dr1,dr2},
although we have also checked their validity
in Pauli-Villars regularization
\cite{pv}.

In the standard-model extension,
all Lorentz- and CPT-violating effects
are controlled by a set of coefficients
that can be regarded as originating
in an underlying theory at the Planck scale.
For example,
they might be associated with expectation values 
in string theory 
\cite{kps},
and specific nonzero coefficients emerge 
in realistic noncommutative field theories
\cite{chklo}.
Several of these coefficients in different sectors
of the standard-model extension
are now bounded by experiments
involving hadrons 
\cite{kexpt,bexpt,kpcvk,bckp},
protons and neutrons
\cite{ccexpt},
electrons
\cite{eexpt,eexpt2},
photons
\cite{photexpt},
and muons
\cite{muexpt}.
Our results in this work can be used to gain insight into the
relationships among coefficients for Lorentz and CPT violation
as the scale ranges between low and high energies.

A basic tool for studying quantum physics over different scales
is the renormalization group
\cite{rg,jc}.
Here,
we discuss its relevance in the context of
the Lorentz- and CPT-violating standard-model extension.
We use our calculations of the one-loop divergences
to extract the corresponding beta functions
for all the coefficients for Lorentz and CPT violation
in the general QED extension.
Solving the associated set of coupled partial-differential equations 
for the renormalized coefficients yields 
their running as the scale is changed. 
Knowledge of this running offers some insight into the
possible relative sizes of nonzero Lorentz- and CPT-violating effects.

This paper is organized as follows.
Section \ref{secttheoframe} 
provides some basic information about 
the general Lorentz- and CPT-violating QED extension.
Renormalizability of the theory
is considered in section \ref{renability}.
Some general issues are discussed,
following which we present the results 
of our one-loop calculation.
We establish the absence of 
divergent cubic and quartic photon interactions,
present explicit results for 
all divergent radiative corrections to the lagrangian,
and show that the Ward identities are preserved at this order.
Section \ref{betafuncs} begins with a discussion of the
application of the renormalization-group method
in the context of Lorentz and CPT violation.
The one-loop beta functions 
for all parameters in the theory are then derived.
The resulting coupled partial-differential equations 
are solved for the running parameters,
and some implications for experiment are considered.
A summary is provided in section \ref{conc}.
The Feynman rules used in our analysis are presented
in the appendix. 
Throughout this work,
our notation is that of Refs.\ \cite{ck,kle}.

\section{Basics}
\label{secttheoframe}

The lagrangian $\cal L$ of the general 
Lorentz- and CPT-violating QED extension
for a fermion field $\ps$ of mass $m$ 
in 4 spacetime dimensions
can be written as 
\cite{ck}
\bea
{\cal L} &=& \frac{1}{2} i \bar{\ps} \Ga^\mu
\stackrel{\leftrightarrow}{D_\mu} \ps  -  \bar{\ps}
M \ps  - \frac{1}{4} F^{\mu\nu} F_{\mu\nu} 
\nonumber \\ & & 
- \frac {1}{4}
(k_F)_{\ka\la\mu\nu} F^{\ka\la} F^{\mu\nu} + \frac{1}{2}
(k_{AF})^{\ka} \ep_{\ka\la\mu\nu} A^\la F^{\mu\nu},
\label{lagdef}
\eea
where $\Ga^{\nu} = \ga^{\nu} + \Ga^{\nu}_1$
and 
$M = m + M_1$,
with
\beq
\Ga^{\nu}_1 &\equiv&
c^{\mu\nu} \ga_\mu + d^{\mu\nu} \ga_5
\ga_\mu + e^\nu + i f^\nu \ga_5 + \half g^{\la\mu\nu}
\si_{\la\mu},
\nonumber \\
M_1 &\equiv& a_\mu \ga^\mu + b_\mu \ga_5 \ga^\mu +
\half H_{\mu\nu} \si^{\mu\nu}.
\label{gamdef}
\eeq
As usual,
we define the covariant derivative
$D_\mu \equiv \prt_\mu + i q A_\mu$
and the electromagnetic field strength
$F_{\mu\nu} \equiv \prt_\mu A_\nu - \prt_\nu A_\mu$.

In the fermion sector,
the coefficients for Lorentz violation
are
\a, \b, \c, \d, \e, \f, \g, \H.
Of these, 
\a, \b, \e, \f, \g\ govern CPT violation.
The coefficients
\a, \b, \H\ have dimensions of mass,
while 
\c, \d, \e, \f, \g\ are dimensionless.
Both \c\ and \d\ can be taken to be traceless,
while \H\ is antisymmetric and \g\ is antisymmetric
on its first two indices.
In the photon sector,
the coefficients for Lorentz violation are
\kkaf, \kkf.
CPT violation is governed only by \kkaf,
which has dimensions of mass.
The coefficient \kkf\ is dimensionless,
has the symmetry properties of the Riemann tensor,
and is double traceless:
\bea
(k_F)_{\ka\la\mu\nu} &=& (k_F)_{\mu\nu\ka\la} = - (k_F)_{\la\ka\mu\nu}, 
\nonumber\\
&&
\hbox{\hskip -50 pt}
(k_F)_{\ka\la\mu\nu} + (k_F)_{\ka\mu\nu\la} + (k_F)_{\ka\nu\la\mu}=0,
\nonumber \\
{(k_F)_{\mu\nu}}^{\mu\nu} &=& 0.
\eea
The requirement that the lagrangian be hermitian 
implies that all the coefficients for Lorentz violation are real.

In the Lorentz-violating theory \rf{lagdef},
two distinct types of Lorentz transformation are relevant 
\cite{ck}.
The lagrangian \rf{lagdef}
is invariant under observer Lorentz transformations:
rotations and boosts of the observer inertial frame
have no effect on the physics
because both the field operators and the coefficients
for Lorentz violation transform covariantly
and because each term in the lagrangian \rf{lagdef}
is an observer scalar.
These coordinate transformations are distinct 
from rotations and boosts of a particle or localized field configuration 
within a fixed observer inertial frame.
The latter are called particle Lorentz transformations.
They leave unchanged the coefficients for Lorentz violation
and hence change the physics.
The theory \rf{lagdef}
therefore violates particle Lorentz invariance.

Since the coefficients for Lorentz violation
are transformed by an observer Lorentz transformation,
an appropriate boost can make at least some of them large. 
To avoid issues with perturbation theory,
in this work we limit calculations to concordant frames
\cite{kle}:
ones in which the coefficients for Lorentz violation are small
compared to the fermion mass $m$ 
or to the dimensionless charge $q$.
Any frame in which the Earth moves nonrelativistically
is known experimentally to be concordant,
so this restriction offers no practical difficulty
in applying our results.
However,
to maintain generality,
we make no assumptions concerning the 
relative sizes of the different coefficients for Lorentz violation.

The hierarchy of scales between the coefficients for Lorentz violation
and the parameters $m$, $q$
has implications for the structure of 
dominant one-loop Lorentz- and CPT-violating effects.
In particular,
since Lorentz and CPT violation can be assumed small
and since we are interested in leading-order 
Lorentz- and CPT-violating effects,
it suffices for the purposes of this work
to define a one-loop diagram 
as one that contains exactly one closed loop 
and is either zeroth or first order 
in coefficients for Lorentz violation.
All relevant one-loop diagrams are therefore  
${\cal O}(q^2)$ 
and at most linear in the coefficients for Lorentz violation.
Note that it would be invalid to include nonlinear contributions 
from the coefficients for Lorentz violation
without also considering multiloop contributions at high order in $q$,
which could be the same order of magnitude.

Combined with symmetry arguments,
the restriction to linear Lorentz- and CPT-violating effects
enables some strong predictions 
about which terms in the lagrangian \rf{lagdef}
can contribute to the renormalization of any given coefficient.
Since QED preserves C, P, and T invariance,
any Lorentz-violating terms mixing linearly under radiative corrections 
must have identical C, P, and T transformation properties.
Table 1 lists these properties for the field operators
appearing in the lagrangian \rf{lagdef}.
For brevity,
the corresponding coefficients for Lorentz violation 
are listed in the table
rather than the field operators themselves.

\medskip
\begin{tabular}{||c|c|c|c||c|c|c||c||}
\hline
\parbox{3cm}{~~~~~~} & $\rm C$ & $\rm P$ & $\rm T$ &
 $\rm CP$ & $\rm CT$ & $\rm PT$ & $\rm CPT$
\\ \hline \hline
\parbox{3cm}
{\quad $c_{00}$,$(k_F)_{0j0k}$, 
\\ \pt{ccc}$c_{jk}$,$(k_F)_{jklm}$}
  &$+$&$+$&$+$&$+$&$+$&$+$&$+$
\\ \hline
$b_j,g_{j0l},g_{jk0},(k_{AF})_j$
  &$+$&$+$&$-$&$+$&$-$&$-$&$-$
\\ \hline
$b_0,g_{j00},g_{jkl},(k_{AF})_0$
  &$+$&$-$&$+$&$-$&$+$&$-$&$-$
\\ \hline
$c_{0j},c_{j0},(k_F)_{0jkl}$
  &$+$&$-$&$-$&$-$&$-$&$+$&$+$
\\ \hline
$a_0,e_0,f_j$
  &$-$&$+$&$+$&$-$&$-$&$+$&$-$
\\ \hline
$H_{jk},d_{0j},d_{j0}$
  &$-$&$+$&$-$&$-$&$+$&$-$&$+$
\\ \hline
$H_{0j},d_{00},d_{jk}$
  &$-$&$-$&$+$&$+$&$-$&$-$&$+$
\\ \hline
$a_j,e_j,f_0$
  &$-$&$-$&$-$&$+$&$+$&$+$&$-$
\\ \hline
\end{tabular}

\medskip

\centerline{\small
Table 1: Discrete-symmetry properties.
}
\smallskip 

The table reveals terms for which the C, P, T symmetries allow mixing 
under renormalization group flow.
Other restrictions also exist.
Since in what follows we adopt a mass-independent renormalization scheme,
operators associated with dimensionless coefficients cannot 
receive corrections from ones associated with massive coefficients. 
Thus,
for example,
$a_\mu$ can receive corrections from $e_\mu$ on symmetry grounds,
but $e_\mu$ cannot receive corrections from $a_\mu$ on dimensional grounds.
There are also restrictions arising from the
rotational invariance of QED.
For instance,
rotational symmetry prevents $e_0$ and $f_j$ from mixing at
this level of approximation,
even though this would be allowed by the C, P, T properties
of the corresponding field operators.
All these features are confirmed by the explicit calculations
that follow.

\section{Renormalizability at one loop}
\label{renability}

In this section,
we give an explicit demonstration of renormalizability at one loop
for the QED extension \rf{lagdef}.
Following some general considerations,
we obtain a generalization of the Furry theorem
and establish the finiteness of the photon vertices.
The divergent propagator and vertex corrections are given
along with the renormalization factors,
and the Ward identities are shown to hold.

\subsection{Setup}

Renormalizability of a quantum field theory 
can be viewed as the requirements 
that the number of primitively divergent 
one-particle-irreducible (1PI) diagrams 
is finite and that the number of parameters 
suffices to absorb the corresponding infinities.
To establish renormalizability of the QED extension \rf{lagdef},
we first determine the superficial degree of divergence 
of a general Feynman diagram.
Using the Feynman rules for the theory 
provided in the appendix,
it follows that the superficial degree of divergence $D$
of a general diagram in the QED extension is
\beq
D &=& 4 - \frac{3}{2} E_\ps - E_A - V_{M_1} - V_{AF},
\label{degdiv}
\eeq
where $E_\ps$ is the number of external fermion legs,
$E_A$ is the number of external photon legs,
$V_{M_1}$ is the number of insertions 
of the $M_1$ operator in a fermion propagator,
and $V_{AF}$ is the number of insertions 
of \kkaf\ in a photon propagator.

The expression \rf{degdiv} shows
that there are a finite number 
of potentially divergent 1PI diagrams at one loop.
Their topologies correspond
to those of the divergent diagrams associated with conventional QED,
displayed in Fig.\ 1.
However,
in addition to these usual diagrams,
there is a set of diagrams obtained from them 
by single insertions of coefficients for Lorentz violation 
allowed by the Feynman rules.
All such insertions lead to 1PI divergent diagrams,
except for those involving the coefficients \a, \b, \H, \kkaf\
inserted into logarithmically divergent diagrams of conventional QED.
These exceptions arise because 
the mass dimensionality of the coefficients 
involved ensures a finite result.

\begin{figure}
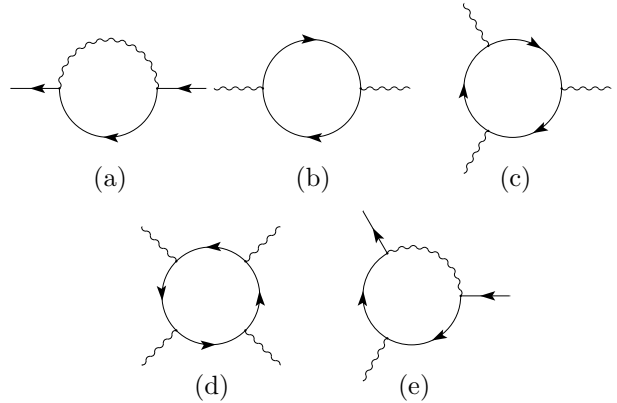

\centering
\begin{tabular}{ccc}
\fgl{(a)}\pad\qedpic{epqed11.eps}\pad &
\fgl{(b)}\pad\qedpic{p2qed11.eps}\pad &
\fgl{(c)}\pad\qedpic{p3qed11.eps}\pad
\end{tabular}
\begin{tabular}{cc}
\fgl{(d)}\pad\qedpic{p4qed11.eps}\pad &
\fgl{(e)}\pad\qedpic{phepqed11.eps}\pad
\\ ~
\end{tabular}
\caption{One-loop topologies for QED.}
\label{gphs}
\end{figure}

\begin{figure}
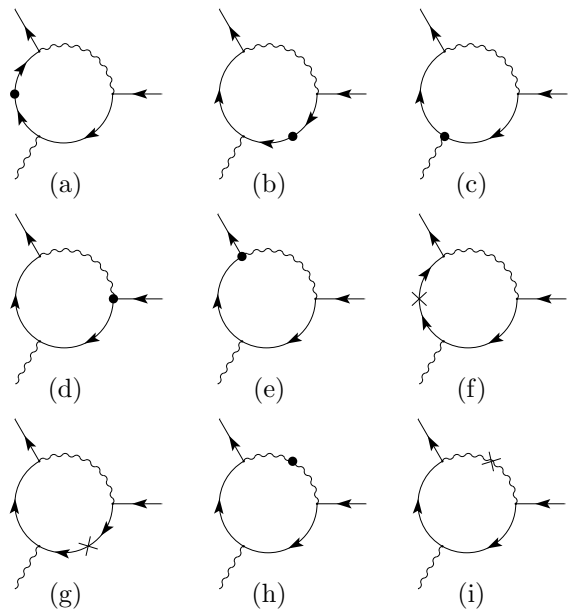

\begin{tabular}{ccc}
\fgl{(a)}\pad \qedpic{phepgam11.eps}\pad
& \fgl{(b)}\pad \qedpic{phepgam12.eps}\pad
& \fgl{(c)}\pad \qedpic{phepgamph11.eps}\pad
\\
\fgl{(d)}\pad \qedpic{phepgamph17.eps}\pad &
\fgl{(e)}\pad \qedpic{phepgamph18.eps}\pad &
\fgl{(f)}\pad \qedpic{phepmass11.eps}\pad
\\
\fgl{(g)}\pad \qedpic{phepmass12.eps}\pad &
\fgl{(h)}\pad \qedpic{phepkayeff11.eps}\pad &
\fgl{(i)}\pad \qedpic{phepkaf11.eps}\pad
\\ ~
\end{tabular}
\caption{Fermion-photon vertices in the QED extension.}
\label{cptgphs}
\end{figure}

As an illustration,
the QED diagram in Fig.\ 1e leads to the set of divergent 
contributions illustrated in Fig.\ 2.
In these diagrams, 
dimensionless coefficients for Lorentz violation 
are represented as filled circles 
while the others are represented by crosses. 
The notation is detailed in the appendix.

As usual,
each additional 1PI diagram involves a one-loop integration.
To evaluate the divergent contributions 
of these diagrams to the effective action,
a regularization scheme for the loop integrations is needed. 
In this paper,
we adopt dimensional regularization.
However,
we have also repeated our calculations 
using Pauli-Villars regularization.
It turns out that the usual correspondence between the two schemes holds,
supporting the expected scheme independence of the physical results.

Certain diagrams involve factors of $\ga_5$,
which introduces complications in dimensional regularization
In dimensional regularization,
the presence of particle Lorentz violation
has little effect on the standard evaluation 
of loop integrals.
Although use is sometimes made of the Lorentz properties of the integrand, 
the standard techniques hold 
because the integrands involve momentum variables
that behave covariantly under both observer and particle transformations,
as usual.
Moreover,
the linearity of Lorentz violation means that
the role of the coefficients for Lorentz violation
is limited in this context to contraction with 
the result of the integration.
For similar reasons,
no new issues arise with manipulations such as Wick rotations.
We can therefore perform the necessary regularization 
of divergent integrals 
by extending spacetime to $d = 4 - 2 \ep$ dimensions
in the conventional way,
so that the one-loop divergent corrections 
to the effective lagrangian take the form of poles
in the infinitesimal parameter $\ep$, 
as usual. 

Certain diagrams involve factors of $\ga_5$,
which introduces complications in dimensional regularization
because the properties of $\ga_5$ are dimension dependent. 
One possibility
is to use the 't Hooft-Veltman definition 
\cite{dr1}
of $\ga_5$,
in which the $\ga$-matrix algebra is infinite-dimensional 
in non-integer dimensions 
and the first four $\ga^\mu$ are treated 
differently from the others. 
In particular,
$\ga_5$ anticommutes with these four 
while commuting with all the others:
\beq
\{ \ga_5, \ga^\mu \} =  0, 
~ \mu \in \{0,1,2,3\},
\quad
[ \ga_5, \ga^\mu ] =  0, 
~ \mu \ge 4. 
\eeq
This procedure introduces a technical breaking of Lorentz invariance 
in all but the first four dimensions,
but without introducing new physical features 
in our perturbation expansion 
because the integrals to be regularized 
have conventional Lorentz properties.
In any case,
in the present context
it is simpler to adopt instead a naive $\ga_5$-matrix
that anticommutes with all of the other $\ga$-matrices,
\beq
\{ \ga_5, \ga^\mu \} &=& 0, \,\,\, \mu \ge 0, 
\eeq
which leads to errors of order $\ep$ in $\ga$-matrix manipulations
and hence to errors in finite terms.
Since we are interested here in the divergences at one loop,
all of which are simple poles in $\ep$, 
the naive $\ga_5$ leaves the poles unaffected while easing calculation.
Determination of the finite radiative contributions at one loop
would require more care but lies beyond our present scope.

Another issue arises because 
the one-loop integrals span an infinite range of four-momentum.
The theory \rf{lagdef} is known  
to violate stability or microcausality at sufficiently high
energy and momentum,
where unrenormalizable terms from Planck-scale physics
become important and must be included in the analysis
\cite{kle}.
The Feynman rules adopted here are therefore strictly valid
only over a range of energy and momentum
lying below the Planck scale.
We proceed in this section under the reasonable 
and customary assumption 
that any new physics entering at high scales 
has negligible effect on the leading-order low-energy physics
described by the lagrangian \rf{lagdef}.
A definitive result concerning the validity of this assumption 
would be of interest.

A technical point to note is that 
no external-leg propagators appear
because we are calculating corrections
to the one-loop effective action.
External legs introduce additional complications
because the full propagator 
at all orders in coefficients for Lorentz violation
is needed to establish the asymptotic Hilbert space
\cite{ck}.
More attention would therefore be required to extract 
the finite radiative corrections to 
physical scattering cross sections or decay amplitudes
at one loop.

We finally remark in passing that,
since Lorentz symmetry is no longer respected by the theory,
certain Lorentz-noninvariant regularization schemes 
might in principle be envisaged instead.
It is conceivable that a scheme chosen to respect
both observer Lorentz invariance and
any remaining subgroup of the particle Lorentz symmetry
might offer some practical or conceptual advantage
over more conventional regularizations.
In any case,
standard dimensional regularization suffices for our purposes here.

\subsection{Generalized Furry theorem \\
and photon-interaction vertices}
\label{cubic}

Renormalizability of the QED extension at one loop
requires that no divergent contributions to the 
three- or four-point photon vertices arise,
since these must be absent in an abelian gauge theory.
In conventional QED, 
the Furry theorem 
\cite{wf} 
plays a useful role in this regard.
In this subsection,
we establish a generalized Furry theorem
and use it and other calculations
to prove the absence of divergent contributions 
to photon interactions at one loop.

In conventional QED,
the Furry theorem relies on the
$\ga$-matrix structure of the photon-fermion vertex,
which leads to a cancellation between two nonzero loops
differing only in the direction of the charge flow. 
However,
the QED extension \rf{lagdef} includes terms
with more general $\ga$-matrix structures.
In this case,
corresponding loops with Lorentz-violating insertions
either cancel or add,
depending on the charge-conjugation properties
of the associated $\ga$-matrix insertion.

\begin{figure}
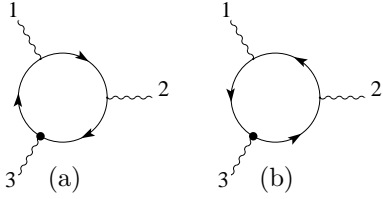

\centering
\begin{tabular}{cc}
\fgl{(a)} \pad \qedpic{p3gamph11.eps}\pad
& \fgl{(b)} \pad \qedpic{p3rgamph11.eps}\pad \\~
\end{tabular}
\caption{Two contributions to the cubic photon interaction.}
\label{cptp3gphs}
\end{figure} 

As an example,
consider the cubic photon vertex at one loop
with an insertion of $\Ga_1^\mu$ 
at one of the fermion-photon vertices in
Fig.\ 1c.
This gives two contributions shown in Fig.\ 3.
Take the loop momentum $k$ to be positive 
in the clockwise direction,
and assign the $n$th external photon line 
a momentum $p_n$ and Lorentz index $\mu_n$.
Define $k_1 = k + p_1$ and $k_{12} = k + p_1 + p_2$.
Then,
the two diagrams yield an expression proportional to
\beq
\int \!\!\! \fr{d^dk}{(2\pi)^d} \left[
\fr{{\rm Tr} [(\fslash{k}+m)\ga^{\mu_1}(\fslash{k}_{1}+m)\ga^{\mu_2}
(\fslash
{k}_{12}+m)\Ga_1^{\mu_3} ]}{(k^2-m^2)(k_1^2-m^2)(k_{12}^2-m^2)} \right.
\nonumber \\ ~ \!\!\!\! - \left.
\fr{{\rm Tr} [\Ga_1^{\mu_3}(\fslash{k}_{12}-m)\ga^{\mu_2}(\fslash{k}_{1}-m)
\ga^{\mu_1}
(\fslash{k}-m) ]}
{(k^2-m^2)(k_1^2-m^2)(k_{12}^2-m^2)} \right]. \!\!\!
\eeq
Taking the transpose of the argument of the trace in the second term
and inserting suitable factors of $C C^{-1}$, 
where $C$ is the charge-conjugation matrix,
we can rewrite the numerator of the integrand as
\beq
{\rm Tr}[(\fslash{k}+m)\ga^{\mu_1}(\fslash{k}_{1}+m)\ga^{\mu_2}(\fslash
{k}_{12}+m) ( \Ga_1^{\mu_3} - \tilde\Ga_1^{\mu_3} ) ],
\eeq
where 
\beq
\tilde\Ga^{\nu}_1 &\equiv&
-(C \Ga^\nu_1 C^{-1} )^T
\nonumber\\
&=& c^{\mu\nu} \ga_\mu - d^{\mu\nu} \ga_5 \ga_\mu 
- e^\nu - i f^\nu \ga_5 
+ \half g^{\la\mu\nu} \si_{\la\mu}.
\eeq
In this case,
it follows that only the terms in $\Ga_1^\mu$
associated with C violation survive:
those involving $I$, $\ga_5$, $\ga_5 \ga^\mu$.
The usual Furry theorem is a special case 
for the C-preserving QED interaction,
with $\Ga_1^\mu$ replaced by $\ga^\mu$.

\begin{figure}
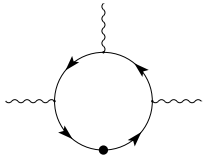

\centering
\qedpic{p3gamprop11.eps}
\caption{Cubic photon vertex with $\Ga_1^\mu$ propagator insertion.}
\label{cubiceee}
\end{figure}     

If instead a factor of $\Ga_1^\mu$ is inserted in a fermion propagator, 
a related argument applies.
See Fig.\ 4.
Three conventional fermion-photon vertices occur,
but an extra fermion propagator appears in the loop
along with another momentum factor from the insertion.
Since the propagator has no net effect 
while the signs from the momentum insertion 
and the extra conventional vertex cancel,
the surviving terms are the same as before. 

For a four-point vertex,
an extra propagator and $\ga$ matrix appear relative 
to the three-point vertex. 
These combine to give an overall relative sign.
It therefore follows that coefficients surviving 
in a three-point function are eliminated
in the corresponding four-point function,
and vice versa.
These arguments can be generalized to include
insertions of $M_1$ in fermion propagators
and arbitrary numbers of photon legs around the loop.

The generalization of the Furry theorem thus shows that
there are no contributions proportional to 
\b, \c, or \g\
for odd numbers of photon legs on a fermion loop,
while there are no contributions proportional to
\a, \d, \e, \f, or \H\
for even numbers of photon legs.
The contributions from other pairs of diagrams
with opposing fermion loops must be explicitly calculated
and typically are nonzero.
This applies to both finite and divergent corrections.
For example,
it is no longer necessarily the case
that one-loop radiative corrections
vanish for $n$-point photon S-matrix amplitudes with odd $n$.
Even for the 3-point photon vertex,
there could now be a nonzero amplitude.
Although it lies outside our present scope,
it would be interesting to evaluate these radiative effects
and consider possible phenomenological implications.

To investigate renormalizability,
it is necessary to calculate explicitly
the divergent one-loop contributions 
to the three- and four-point photon vertices
for those cases where the generalized Furry theorem
allows a nonzero answer.
For the cubic vertex there are three vertices and three propagators,
so any nonzero divergent contribution would occur three times.
The resulting permutations are illustrated in Fig.\ 5
for the case of propagator insertions.
Any diagram with an \H\ insertion is finite 
either because of the dimensionality of \H\ 
or because the trace of an odd number of $\ga$-matrices vanishes.
Explicit calculation reveals that
no divergences proportional to the other 
coefficients for Lorentz violation occur either.

\begin{figure}
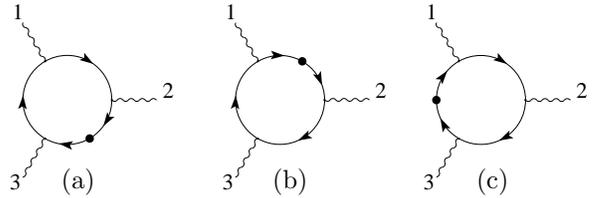

\centering
\begin{tabular}{ccc}
\hspace{-0.5cm}
\fgl{(a)}\pad \qedpic{p3ec1.eps} \pad &
\fgl{(b)}\pad  \qedpic{p3ec2.eps}\pad &
\fgl{(c)}\pad  \qedpic{p3ec3.eps}\pad \\ ~
\end{tabular}
\caption{Permutations for the cubic photon vertex.}
\label{ecgphs}
\end{figure} 

For the quartic photon vertex,
the generalized Furry theorem and the dimensionality of \b\ 
together imply that the only potential divergences 
arise from insertions of \c\ or \g.
In this case,
there are four permutations of each type of insertion.
The contributions involving \g\ are zero because 
the trace of an odd number of $\ga$ matrices vanishes.
Once again,
explicit calculation reveals that no divergence
proportional to \c\ occurs either, 
as required for renormalizability.

\subsection{Propagator corrections}
\label{twopoint}

In this subsection,
we provide the results of our calculations for
one-loop corrections to the photon and fermion propagators.
The calculational methods parallel the conventional case,
so for brevity we restrict the discussion largely to 
the presentation of results.

For the photon propagator,
the complete one-loop divergence
including the standard QED result is
\beq
\bar{\om}_{\mu\nu} (p)
&=& \fr{4q^2}{3} I_0 \Big[ \left( p_\mu p_\nu - p^2
\et_{\mu\nu} \right) \nonumber
 \\
& & \nonumber 
- \left( c_{\mu\nu} + c_{\nu\mu} \right) p^2 
- 2 c_{\al \be} p^\al p^\be \et_{\mu\nu} 
\nonumber \\
& &  
+ \left( c_{\mu\al} + c_{\al\mu} 
\right) p^\al p_\nu 
+ \left( c_{\nu\al} + c_{\al\nu} 
\right) p^\al p_\mu 
\Big],
\nonumber\\
\label{phtwo}
\eeq
where $I_0 = {i}/{16\pi^2 \ep}$.
Only the symmetric part of $c_{\mu\nu}$ contributes. 
Corrections coming from diagrams involving vertex and propagator insertions
of $g^{\la\mu\nu}$ and the antisymmetric part of $c^{\mu\nu}$ cancel.
All other potential divergent corrections to this propagator 
can be shown to vanish,
using arguments similar to those in the previous subsection.
Note that $\bar{\om}_{\mu\nu}$ is symmetric and
gauge invariant, $p^\mu \bar{\om}_{\mu\nu} = 0$.
The result \rf{phtwo} agrees with the original
calculation in Ref.\ \cite{ck}.

For the fermion propagator,
the usual QED correction is
\beq
\Si_{\rm QED}(p) &=& 
q^2 I_0 \Big[ ( 1 + \xi ) \ga_\mu p^\mu - (4+\xi) m
\Big].
\label{qedtwo}
\eeq
The divergent corrections $\Si_x$
arising from all possible insertions of an $x$ term 
in the standard one-loop QED diagram
are given by
\beq
\Si_{M_1}(p) &=&  - q^2 I_0 \Big[ (1+\xi) (a_\mu + \ga_5 b_\mu) \ga^\mu
+ \fr{\xi}{2} H_{\mu\nu} \si^{\mu\nu} \Big],
\nonumber\\ 
\Si_c(p) &=& \fr{q^2}{3} I_0 
\Big[ ( 3\xi - 1) c_{\nu\mu} p^\mu \ga^\nu 
- 4 c_{\mu\nu} p^\mu \ga^\nu \Big],
\nonumber\\
\Si_d(p) &=& \fr{q^2}{3} I_0 \Big[ ( 3\xi - 1) 
d_{\nu\mu} p^\mu \ga_5 \ga^\nu 
- 4 d_{\mu\nu} p^\mu \ga_5 \ga^\nu
\nonumber\\
& & 
\qquad \qquad
- 3  m \ep^{\al\be\mu\nu} d_{\al\be} \si_{\mu\nu} \Big],
\nonumber\\
\Si_e(p) &=& q^2 I_0 \Big[ ( 1 + \xi ) e_\mu p^\mu - 3 m e_\mu \ga^\mu \Big],
\nonumber\\
\Si_f(p) &=& q^2 I_0 ( 1 + \xi ) i f_\mu p^\mu \ga_5,
\nonumber\\
\Si_g(p) &=& 
\fr{q^2}{2} I_0 \Big[ ( \xi - 1) g_{\la\mu\nu} p^\nu \si^{\la\mu} 
- 2 g_{\la\mu\nu} p^\mu \si^{\la\nu}  
\nonumber\\
& & 
\qquad \qquad
+ 2 g_{\mu\al}^{\pt{\al\mu}\al} p_\nu \si^{\mu\nu} 
-  m g^{\al\be\ga} \ep_{\al\be\ga\mu} \ga_5 \ga^\mu \Big],
\nonumber\\
\Si_{k_{AF}}(p) &=& 3 q^2 I_0  (k_{AF})_\nu \ga_5 \ga^\nu ,
\nonumber\\
\Si_{k_F}(p) &=&  \fr{ 4 q^2}{3} I_0  
(k_F)_{\mu\al\nu}^{{\pt{\mu\la\mu}}\al}
p^\mu \ga^\nu.
\label{stfp}
\eeq

\subsection{Quadratic-term renormalization factors}

To renormalize the quadratic terms,
we must redefine the bare fields and the fermion mass 
in terms of renormalized ones,
\beq
\ps_B &=& \sqrt{Z_\ps} \ps,\quad 
A^\mu_B = \sqrt{Z_A} A^\mu,\quad
m_B = Z_m m,
\eeq
and the bare coefficients for Lorentz violation 
in terms of renormalized ones,
\bea
a_{B\mu} = 
(Z_a)_\mu^{\pt{\mu}\al} a_\al,
&\quad&
b_{B\mu} = 
(Z_b)_\mu^{\pt{\mu}\al} b_\al,
\nonumber\\
c_{B\mu\nu} = 
(Z_c)_{\mu\nu}^{\pt{\mu\nu}\al\be} c_{\al\be},
&\quad&
d_{B\mu\nu} = 
(Z_d)_{\mu\nu}^{\pt{\mu\nu}\al\be} d_{\al\be},
\nonumber\\
e_{B\mu} = 
(Z_e)_\mu^{\pt{\mu}\al} e_\al,
&\quad&
f_{B\mu} = 
(Z_f)_\mu^{\pt{\mu}\al} f_\al,
\nonumber\\
g_{B\la\mu\nu} = 
(Z_g)_{\la\mu\nu}^{\pt{\la\mu\nu}\al\be\ga} g_{\al\be\ga},
&\quad&
H_{B\mu\nu} = 
(Z_H)_{\mu\nu}^{\pt{\mu\nu}\al\be} H_{\al\be},
\nonumber\\
&&\hbox{\hskip -80pt}
(k_{AF})_{B\mu} = 
(Z_{k_{AF}})_\mu^{\pt{\mu}\al} (k_{AF})_\al,
\nonumber\\
&&\hbox{\hskip -80pt}
(k_F)_{B\ka\la\mu\nu} = 
(Z_{k_F})_{\ka\la\mu\nu}^{\pt{\ka\la\mu\nu}\al\be\ga\de}
(k_F)_{\al\be\ga\de}.
\label{bare}
\eea
In this subsection,
a subscript $B$ is added to bare quantities
where needed to distinguish them from renormalized ones.

An analysis of Eqs.\ \rf{qedtwo} and \rf{stfp}
leads to the following expressions 
for the above renormalization factors:
\beq
&
Z_\ps = 1 - \fr{q^2}{16\pi^2 \ep} (1+\xi), 
\quad
Z_A = 1 - \fr{q^2}{12\pi^2 \ep}, 
&
\nonumber\\
&
Z_m = 1 - \fr{3 q^2}{16\pi^2 \ep}, 
&
\nonumber\\
&
(Z_a)_\mu^{\pt{\mu}\al} a_\al 
= a_\mu - \fr{3 q^2}{16\pi^2 \ep} m e_\mu, 
&
\nonumber\\
&
(Z_b)_\mu^{\pt{\mu}\al} b_\al
= b_\mu - \fr{q^2}{32\pi^2 \ep}
(m g^{\al\be\ga} \ep_{\al\be\ga\mu} - 6 (k_{AF})_\mu), 
&
\nonumber\\
&
(Z_c)_{\mu\nu}^{\pt{\mu\nu}\al\be} c_{\al\be}
= c_{\mu\nu} + \fr{q^2}{12\pi^2 \ep}
( c_{\mu\nu} + c_{\nu\mu} - (k_F)_{\mu\la\nu}^{{\pt{\mu\la\mu}}\la}),
&
\nonumber\\
&
(Z_d)_{\mu\nu}^{\pt{\mu\nu}\al\be} d_{\al\be}
= d_{\mu\nu} + \fr{q^2}{12\pi^2 \ep}
( d_{\mu\nu} + d_{\nu\mu} ),  
&
\nonumber\\
&
(Z_e)_{\mu\nu}
=(Z_f)_{\mu\nu}
= \et_{\mu\nu},  
&
\nonumber\\
&
\hbox{\hskip -30 pt}
(Z_g)_{\la\mu\nu}^{\pt{\la\mu\nu}\al\be\ga} g_{\al\be\ga}
= g_{\la\mu\nu} +   
\fr{q^2}{16\pi^2 \ep}  
(2g_{\la\mu\nu} 
&
\nonumber\\
&
\hbox{\hskip 55 pt}
- g_{\nu\la\mu}
+ g_{\nu\mu\la} 
- \et_{\mu\nu} g_{\la\de}^{\pt{\la]\de}\de} 
+ \et_{\la\nu} g_{\mu\de}^{\pt{\la]\de}\de} 
),
&
\nonumber\\
&
\hbox{\hskip -100 pt}
(Z_H)_{\mu\nu}^{\pt{\mu\nu}\al\be} H_{\al\be}
= H_{\mu\nu}
&
\nonumber\\
&
\hbox{\hskip 80 pt}
+ \fr{q^2}{16\pi^2 \ep}
( H_{\mu\nu} - 2 m d^{\al\be} \ep_{\al\be\mu\nu} ), 
&
\nonumber\\
&
(Z_{k_{AF}})_\mu^{\pt{\mu}\al} (k_{AF})_\al
= (k_{AF})_\mu + \fr{q^2}{12\pi^2 \ep} (k_{AF})_\mu,
&
\nonumber\\
& 
\hbox{\hskip -100 pt}
(Z_{k_F})_{\ka\la\mu\nu}^{\pt{\ka\la\mu\nu}\al\be\ga\de}
(k_F)_{\al\be\ga\de}
&
\nonumber\\ 
&
\hbox{\hskip -30 pt}
= (k_F)_{\ka\la\mu\nu} 
+ \fr{q^2}{12\pi^2 \ep} 
\Big[ 
(k_F)_{\ka\la\mu\nu}
&
\nonumber\\ 
&
\hbox{\hskip 50 pt}
- \half
\et_{\mu\ka} ( c_{\nu\la} + c_{\la\nu})
+ \half
\et_{\nu\ka} ( c_{\mu\la} + c_{\la\mu}) 
\pt{\Big],}
&
\nonumber\\ 
&
\hbox{\hskip 50 pt}
+ \half
\et_{\mu\la} ( c_{\nu\ka} + c_{\ka\nu}) 
- \half
\et_{\nu\la} ( c_{\mu\ka} + c_{\ka\mu})
\Big].
&
\label{zfall}
\eeq
We find that these renormalization factors
suffice to render finite at one loop all corrections 
to the quadratic fermion and photon terms.

The derivation of Eqs.\ \rf{zfall}
parallels the standard QED case.
For example,
the only correction to the Lorentz- and CPT-invariant 
fermion kinetic term 
comes from $\Si_{\rm QED}(p)$, as usual.
We therefore have
\bea
&
\hbox{\hskip -50pt}
i \bar{\ps_B} \fslash{p} \ps_B +
\bar{\ps_B} \Si_{\rm QED}(p)|_{m=0} \ps_B + \ldots 
&
\nonumber \\
&
\hbox{\hskip 30pt}
=i Z_\ps \bar{\ps} \fslash{p}\ps 
+ q^2 I_0 \bar{\ps} ( 1 + \xi ) \fslash{p} \ps + \ldots,
&
\eea
where the right-hand side is written 
in terms of one-loop renormalized quantities 
and the ellipsis refers to higher-order terms that can be neglected.
The right-hand side of this equation must be finite,
yielding the first of Eqs.\ \rf{zfall}.

For the coefficients for Lorentz violation,
similar methods apply.
For example,
for the $e_\mu$ term 
we find that the only relevant corrections come from 
$\Si_e(p)|_{m=0}$.
Including the effects arising from 
the wave-function renormalization $Z_\ps$
reveals that no renormalization of 
$e^\mu$ is needed at one loop.
A more involved example
is given by the corrections to $c_{\mu\nu}$,
which arise from both $\Si_c(p)$ and $\Si_{k_F}(p)$.
Incorporating also the wave-function renormalization $Z_\ps$
leads to the above expression for 
$(Z_c)_{\mu\nu}^{\pt{\mu\nu}\al\be} c_{\al\be}$.
As a final example,
consider $(Z_{k_F})_{\ka\la\mu\nu}^{\pt{\ka\la\mu\nu}\al\be\ga\de}$.
Here,
it is useful to note that 
the term $- 2 p^\al p^\be c_{\al \be} \et_{\mu\nu}$ 
in Eq.\ \rf{phtwo} 
must cancel a correction 
to the tree-level $(k_F)_{\mu\nu\al\be}$ term,
and hence the correction itself must also have the
symmetry of the Riemann tensor.
Implementing this requirement reveals that all the divergent terms
cancel simultaneously
provided
$Z_A$ takes the same form as in conventional QED at one loop.

\subsection{Vertex corrections and Ward identities}
\label{wardids}

The remaining 1PI diagrams arise in connection with the
one-loop three-point vertex.
This subsection presents the results of our calculations of 
the associated diagrams.

For this vertex,
the standard QED divergence at one loop is recovered:
\beq
\Ga_{\rm QED}^\mu = - q^3 I_0 ( 1 + \xi ) \ga^\mu.
\label{qedvertex}
\eeq
The divergent corrections $\Ga_x^\mu$
arising from all possible single insertions of an $x$ term
in the standard one-loop QED three-point vertex 
are found to be
\bea
\Ga_{M_1}^\mu &=&  0,
\nonumber\\
\Ga_c^\mu &=& - \fr{q^3}{3} I_0 \Big[ ( 3\xi - 1)  c^{\al\mu} \ga_\al 
- 4 c^{\mu\al} \ga_\al \Big],
\nonumber\\
\Ga_d^\mu &=& - \fr{q^3}{3} I_0 \Big[ ( 3\xi - 1) d^{\al\mu} \ga_5 \ga_\al 
- 4 d^{\mu\al} \ga_5 \ga_\al \Big],
\nonumber\\
\Ga_e^\mu &=& - q^3 I_0 ( 1 + \xi ) e^\mu,  
\nonumber\\
\Ga_f^\mu &=& - q^3 I_0 ( 1 + \xi ) i f^\mu  \ga_5,  
\nonumber\\
\Ga_g^\mu &=& - \fr{q^3}{2} I_0 \Big[ ( \xi - 1) g^{\al\be\mu} \si_{\al\be} 
+ 2 g_{{\pt{\al}}\be}^{\al{\pt{\be}}\be} \si_{\al\mu} 
\nonumber \\
& & 
\qquad\qquad
+  2 g^{\al\mu\be} \si_{\be\al} \Big], 
\nonumber\\
\Ga_{k_{AF}}^\mu &=& 0, 
\nonumber\\
\Ga_{k_F}^\mu &=&  - \fr{ 4 q^3}{3} I_0 
{(k_F)_{\mu\al\be}}^\al \ga^\be.
\label{vertwo}
\eea
The reader is cautioned not to confuse 
the standard notation $\Ga_x^\mu$ 
for the divergent 3-vertex contributions 
with the quantities $\Ga^\mu$ and $\Ga^\mu_1$
in Eqs.\ \rf{lagdef} and \rf{gamdef}.

Taking into account the results \rf{zfall},
renormalization of the QED vertex correction \rf{qedvertex} yields
\beq
q_B = Z_q q, \quad && Z_q = 1 + \fr{q^2}{24\pi^2 \ep}.
\label{zfacque}
\eeq
This is in accordance with the usual QED Ward identity,
\beq
Z_q\sqrt{Z_A} = 1.
\eeq
Note that the $Z$ factors for the couplings 
are all independent of the gauge parameter
$\al = ( 1 - \xi )$,
as expected.

At this stage no more parameters can be renormalized,
so all the remaining divergent corrections $\Ga_x^\mu$ 
must be made finite by the $Z$ factors already defined.
Inspection shows that this is indeed the case,
provided the Ward identity holds.
We can therefore conclude that
the theory is multiplicatively renormalizable
and that it remains gauge invariant at one loop.

\section{Renormalization group \\
and beta functions}
\label{betafuncs}

Given the results for the one-loop divergences presented 
in the previous section,
several interesting issues become amenable to analysis.
In particular,
we can initiate a study 
of the behavior of the Lorentz- and CPT-violating QED extension
over a large range of energy scales.
This offers,
for example,
partial insight about the absolute and relative sizes 
of the coefficients for Lorentz violation
and hence is of value from both theoretical and experimental perspectives.

In this section, 
we begin with some preliminary remarks delineating
the framework for our analysis.
We then obtain the beta functions 
and solve the renormalization-group equations
for the running of the coefficients for Lorentz violation.
Some implications of these results are discussed.

\subsection{Framework}

Any regularization scheme naturally introduces a mass scale $\mu$.
In the Pauli-Villars scheme
it is the momentum cutoff,
while in dimensional regularization it is the unit of mass 
required to maintain a dimensionless action in $d$ dimensions.
In a conventional field theory,
the bare Green functions are independent 
of the value chosen for $\mu$.
However,
they are related by multiplicative renormalizability 
to the renormalized Green functions,
which may depend on $\mu$ 
as a consequence of the regularization procedure.

In perturbation theory,
the coefficients of a renormalized Green function 
typically depend on logarithms $\ln (p^2/\mu^2)$
of the momentum $p$.
As a result,
if $\sqrt{p^2}$ is very different from $\mu$,
perturbation theory can become invalid 
even for small couplings.
To study the physics at momenta much larger than $\mu$,
the Green function must be expressed in terms 
of a new renormalization mass $\mu' \approx \sqrt{p^2}$
chosen to keep the logarithms small,
so that it is justifiable to neglect higher-order terms.

In the present case,
we are interested in the behavior 
of the coefficients for Lorentz violation 
over a large range of scales.
We are therefore interested in the dependence  
of the renormalized Green functions 
on the renormalization mass scale $\mu$.
A standard procedure to obtain this is 
to solve the renormalization-group equation
at the appropriate order in perturbation theory
\cite{rg,jc}.
The solution can be expressed through the dependence on $\mu$
of the renormalized running couplings $x(\mu)$,
where $x$ generically denotes parameters in the theory.

Under the assumption that the Lorentz- and CPT-violating QED extension 
is a multiplicatively renormalizable theory to all orders,
the usual derivation of the renormalization-group equation
can be followed.
For an $n$-point Green function $\Ga(n)$,
we can write
\beq
\Ga_B(n)[p, x_B] &=& Z^{-n/2}[x(\mu), \mu] \Ga(n)[p, x(\mu), \mu],
\label{rge}
\eeq
where $x$ includes the coefficients for Lorentz violation
as well as $m$ and $q$.
The factor of $Z^{-n/2}$ arises from wave-function renormalization 
of the external legs of the Green function.
Noting that the left-hand side of this equation is independent of $\mu$,
differentiation with respect to $\mu$ yields
\beq
0 &=& \mu \fr{d}{d\mu} \left\{ Z^{-n/2}[x(\mu), \mu] \Ga(n)[p, x(\mu), \mu]
\right\}.
\label{rgeq}
\eeq
This is the renormalization-group equation,
which provides a nontrivial constraint on $\Ga(n)$ 
through the explicit $\mu$-dependence.

For a one-loop calculation,
it suffices to impose Eq.\ \rf{rgeq} only to one-loop order.
If the running couplings become large enough,
the perturbative approach fails.
However,
in the region where the couplings remain small,
the accuracy of the perturbative approximation is improved
compared to using fixed couplings.
This improvement can be attributed 
to a partial resummation of the perturbation series 
that includes leading logarithmic corrections
to all orders in perturbation theory.

Despite the link between higher loops 
and renormalization-group improvement,
the explicit one-loop solution of Eq.\ \rf{rgeq}
only uses multiplicative renormalizability at one loop.
Since one-loop multiplicative renormalizability 
of the Lorentz- and CPT-violating QED extension 
is proved in the previous section,
it follows that we can perform
one-loop renormalization-group calculations
without meeting practical obstacles.
In effect,
this procedure makes the reasonable assumption
that the couplings remain small
and hence that the perturbation approximation is valid.
However,
the derivation of the renormalization-group equation outlined above
shows that adopting this calculational procedure
also tacitly assumes that the QED extension 
is multiplicatively renormalizable at all orders.
Resolving this multiloop issue lies well beyond our present scope,
so we restrict ourselves here to a few pertinent remarks.

It is known that operators of mass dimension greater than four
are required in the full QED extension for causality and stability 
at Planck-related energy scales
\cite{kle}.
Although such operators are unnecessary at low energy,
where the theory \rf{lagdef} holds sway,
their presence implies that the full theory 
is unrenormalizable in the usual sense.
However,
this may be irrelevant in the underlying Planck-scale theory.
For example,
difficulties with ultraviolet properties are absent
in some string theories
even though the corresponding particle field theories
appear to have unrenormalizable terms at leading order
in the string tension.
In the present context,
the existence of these operators
suggests that the lagrangian \rf{lagdef}
can be regarded as an effective low-energy theory.
From this perspective,
the key assumption made in applying Eq.\ \rf{rgeq} at one loop
is that the theory remains valid 
as an effective field theory at this order.
Then,
any problems arising from unrenormalizability at higher loops 
would be suppressed at low energies,
allowing one-loop calculations to be performed as though 
the theory were fully multiplicatively renormalizable.
Unrenormalizable effects entering at some multiloop level
might cause loss of predictability 
at that order in perturbation theory,
but lower-order predictions would remain valid
in the region where the corresponding running couplings are small.

For definiteness,
we proceed in what follows under the reasonable and practical assumption 
that it is meaningful to apply Eq.\ \rf{rgeq} at one loop.
Although beyond our present scope,
it would be interesting either to prove all-orders multiplicative
renormalizability of the QED extension
or to determine the formal regime of validity of the results
to be presented below.  

\subsection{The beta functions}

Given a theory with a set 
$\{x_j\}, j = 1,2,\ldots,N$
of running parameters,
the beta function 
\cite{jc}
for a specific parameter $x_j$ is 
\beq
\be_{x_j} &\equiv& \mu \fr{d x_j}{d\mu},
\label{betadef}
\eeq
where $\mu$ is the renormalization mass parameter 
relevant to the regularization method.
In dimensional regularization with minimal subtraction,
the beta function for a given parameter 
can be calculated directly from the simple $\ep$ pole 
in the corresponding $Z$ factor 
\cite{gt,betfun}.
In this subsection,
we summarize the procedure
and use it to obtain all the one-loop beta functions for
the Lorentz- and CPT-violating QED extension.

For each parameter $x_j$,
define the $Z$ factor $Z_{x_j}$ by
\beq
x_{jB} &=& \mu^{\rh_{x_j} \ep} Z_{x_j} x_j,
\label{baredef}
\eeq
where the factor $\mu^{\rh_{x_i}}$ 
ensures that the renormalized parameter $x_j$
has the same mass dimension as its corresponding bare parameter.
In $d = 4 - 2 \ep$ dimensions,
the fields $\ps$ and $A^\mu$ have mass dimension 
$(3-2\ep)/2$ and 
$(1-\ep)$,
respectively.
We therefore find
\beq
\rh_q &=& 1,
\quad
\rh_m = \rh_{\Ga_1} = \rh_{M_1} = 0.
\eeq

In minimal subtraction,
only the divergent terms in the regulated integrals
are subtracted.
Since these divergent terms are poles in $\ep$,
any given $Z$ factor takes the generic form
\beq
Z_{x_j} x_j &=& x_j + \sum_{n=1}^\infty \fr{a^j_n}{\ep^n}.
\label{zeddef}
\eeq
It can then be shown that
\cite{gt}
\beq
\be_{x_j} &=& 
\lim_{\ep \to 0}\Big[ 
- \rh_{x_j} a^j_1 
+ \sum_{k=1}^{N} \rh_{x_k} x_k \fr{\prt a^j_1 }{\prt x_k} 
\Big],
\label{betacalc}
\eeq
which involves only the simple pole $a_1^j$.

Among the $\rh_{x_j}$,
only $\rh_q$ is nonzero in the present case.
Equation \rf{betacalc} therefore implies that
each beta function other than $\be_q$ is nontrivial only  
due to the $q$ dependence of the corresponding $Z$ factor.
The running of all the coefficients for Lorentz violation
at any loop order
is therefore driven by the standard QED running of the charge $q$.
This result is expected 
because the interactions in the theory \rf{lagdef} 
all arise from the minimal coupling in the covariant derivative
and are therefore controlled only by the charge $q$,
even though the perturbation theory is also an expansion
in the coefficients for Lorentz violation.

Using the expression \rf{betacalc},
we obtain the following set of results for all the beta functions
in the Lorentz- and CPT-violating QED extension:
\beq
\be_{m} &=& - \fr{3 q^2}{8\pi^2}m,
\quad
\be_{q} = \fr{q^3}{12\pi^2}, 
\nonumber\\
(\be_a)_{\mu} &=& - \fr{3q^2}{8\pi^2} m e_\mu, 
\nonumber\\
(\be_b)_{\mu} &=& - \fr{q^2}{16\pi^2} 
( m g^{\al\be\ga} \ep_{\al\be\ga\mu} - 6 \kaf_\mu ),
\nonumber\\
(\be_c)_{\mu\nu} &=& \fr{q^2}{6\pi^2} ( c_{\mu\nu} + c_{\nu\mu} -
\kf_{\mu\al\nu}^{\pt{\mu\al\nu}\al} ), 
\nonumber\\
(\be_d)_{\mu\nu} &=& \fr{q^2}{6\pi^2} ( d_{\mu\nu} + d_{\nu\mu} ),
\quad 
(\be_e)_{\mu} =
(\be_f){_\mu}= 0, 
\nonumber\\
(\be_g)_{\la\mu\nu} &=& \fr{q^2}{8\pi^2} 
( 
2g_{\la\mu\nu}
- g_{\nu\la\mu} + g_{\nu\mu\la}
\nonumber\\
&&\qquad\qquad
- \et_{\mu\nu} g_{\la\al}^{\pt{\mu\al}\al} 
+ \et_{\la\nu} g_{\mu\al}^{\pt{\nu\al}\al} 
), 
\nonumber\\
(\be_H)_{\mu\nu} &=&  \fr{q^2}{8\pi^2} 
( H_{\mu\nu} - 2 m d^{\al\be} \ep_{\al\be\mu\nu}), 
\nonumber\\
(\be_{k_{AF}})_{\mu} &=&  \fr{q^2}{6\pi^2} \kaf_\mu, 
\nonumber\\
(\be_{k_F})_{\ka\la\mu\nu}
&=& \fr{q^2}{6\pi^2} 
\Big[ 
{k_F}_{\ka\la\mu\nu} 
\nonumber\\
&&\qquad
- \half
\et_{\mu\ka} ( c_{\nu\la} + c_{\la\nu})
+ \half
\et_{\nu\ka} ( c_{\mu\la} + c_{\la\mu}) 
\nonumber\\
&&\qquad
+ \half
\et_{\mu\la} ( c_{\nu\ka} + c_{\ka\nu}) 
- \half
\et_{\nu\la} ( c_{\mu\ka} + c_{\ka\mu})
\Big]. 
\nonumber\\
\label{betall}
\eeq
Through the definition \rf{betadef},
these beta functions specify a complete set 
of coupled partial-differential equations 
governing the one-loop running of the coefficients
for Lorentz violation and other parameters
in the theory.

\subsection{Running couplings}
\label{running}

To solve the coupled partial-differential equations \rf{betall}
for the running parameters $x_j(\mu)$,
boundary conditions are required.
We provide these as the values 
\beq
x_{j0} &\equiv& x_j(\mu_0)
\eeq
of the parameters $x_j$ at the scale $\mu_0$.
In what follows,
it is convenient to define the quantity 
\beq
Q(\mu) &\equiv& 1 - \fr{q_0^2}{6\pi^2} \ln \fr{\mu}{\mu_0},
\label{qdef}
\eeq
which controls the running of the usual QED charge
with the scale $\mu$.

The first two of Eqs.\ \rf{betall} yield 
\beq
q(\mu)^2 &=& Q^{-1} q_0^2,
\quad m(\mu) = Q^{9/4} m_0,
\label{mqrun}
\eeq
which is the standard QED result. 
Substituting these expressions in the remaining equations 
permits a complete integration.
As an example,
consider the coefficient $d_{\mu\nu}$.
Constructing the beta function for $d_{\mu\nu} + d_{\nu\mu}$ gives
\beq
\fr{d}{d\ln \mu} (d + d^T) &=& \fr{q_0^2}{3\pi^2} Q^{-1} ( d + d^T),
\eeq
which can be rewritten as
\beq
\fr{d}{d\ln \mu} \Big[ Q^2 (d + d^T) \Big] &=& 0.
\eeq
Therefore,
the symmetric part of $d_{\mu\nu}$ runs like $Q^{-2}$.
Similarly,
we find the antisymmetric part has no running.
Combining the two gives the running of $d_{\mu\nu}$.

In this way,
we find that the coefficients for Lorentz violation
run as follows:
\beq
a_\mu &=& a_{0\mu} - m_0(1 - Q^{9/4}) e_{0\mu},
\nonumber\\
b_\mu &=& b_{0\mu} - 
\frac 16 m_0(1 - Q^{9/4}) g^{\al\be\ga}_0 \ep_{\al\be\ga\mu}
\nonumber\\
&&\qquad
- \frac 94 \ln Q ~\kaf_{0\mu},
\nonumber\\
c_{\mu\nu} &=& 
c_{0\mu\nu} 
\nonumber\\ &&
- \frac 13 (1 - Q^{-3})
( c_{0\mu\nu} + c_{0\nu\mu} 
- \kf_{0\mu\al\nu}^{\pt{0\mu\al\nu}\al} ) ,
\nonumber\\
d_{\mu\nu} &=& 
d_{0\mu\nu}
- \half (1 - Q^{-2}) ( d_{0\mu\nu} + d_{0\nu\mu} ) ,
\nonumber\\
e_\mu &=& e_{0\mu},
\quad 
f_\mu = f_{0\mu},
\nonumber\\
g_{\la\mu\nu} &=&
g_{0\la\mu\nu} 
- \frac 13 (1 - Q^{-9/4})
\nonumber\\  &&
\qquad 
\times
(2 g_{0\la\mu\nu} 
- g_{0\nu\la\mu} + g_{0\nu\mu\la} 
\nonumber\\
&&\qquad \qquad
- \et_{\mu\nu} g_{0\la\al}^{\pt{0\la\al}\al}
+ \et_{\nu\la} g_{0\mu\al}^{\pt{0\la\al}\al} ) ,
\nonumber \\
H_{\mu\nu} &=& 
H_{0\mu\nu} + \half m_0 (1 - Q^{9/4}) d_0^{\al\be} \ep_{\al\be\mu\nu} ,
\nonumber\\
\kaf_\mu &=& Q^{-1} \kaf_{0\mu},
\nonumber\\
\kf_{\ka\la\mu\nu} 
&=& \kf_{0\ka\la\mu\nu} 
+ \frac 16 ( 1 - Q^{-3})
\nonumber\\
&&\qquad
\times\Big[
\et_{\mu\ka} ( c_{\nu\la} + c_{\la\nu}
- \kf_{0\nu\al\la}^{\pt{0\mu\al\nu}\al} ) 
\nonumber\\
&&\qquad
- \et_{\nu\ka} ( c_{\mu\la} + c_{\la\mu} 
+ \kf_{0\mu\al\la}^{\pt{0\mu\al\nu}\al} ) 
\nonumber\\
&&\qquad
- \et_{\mu\la} ( c_{\nu\ka} + c_{\ka\nu}
+ \kf_{0\nu\al\ka}^{\pt{0\mu\al\nu}\al} ) 
\nonumber\\
&&\qquad
+ \et_{\nu\la} ( c_{\mu\ka} + c_{\ka\mu}
- \kf_{0\mu\al\ka}^{\pt{0\mu\al\nu}\al} ) 
\Big].
\label{runtwo}
\eeq
Some coefficients increase or decrease with $Q$
while some are unaffected,
including irreducible combinations such as
the totally antisymmetric part of \g.
Note that
the mixings of coefficients for Lorentz violation
displayed in these results 
are consistent with the symmetry-based predictions
given at the end of section \ref{secttheoframe}.

\subsection{Some implications}

The expressions \rf{runtwo}
display a range of behavior for the
running of the coefficients for Lorentz violation.
All the running is controlled by the single function $Q(\mu)$ 
given in Eq.\ \eqref{qdef},
but the powers of $Q(\mu)$ involved range from $-3$ to $9/4$.  
In this subsection,
we comment on some implications of this behavior.

Note first that
the calculations above are performed
for the Lorentz- and CPT-violating QED extension
with a single Dirac fermion.
However,
the full standard-model extension involves 
additional interactions,
three generations of chiral fermion multiplets,
a scalar field,
and several intermeshed sets of coefficients for Lorentz violation.
All these would affect the structure of the 
solutions \rf{runtwo}.
A definitive understanding of the physical implications
of the running of the coefficients for Lorentz violation
must therefore await a complete analysis within the 
standard-model extension.
Nonetheless,
despite these caveats,
some meaningful physical insight can be obtained.

We focus here on the behavior of $Q(\mu)$
from low energies to the Planck scale.
A key factor in determining this behavior 
is the value of the fermion charge $q_0$
at the reference scale $\mu_0$,
which controls the size of the coefficient
of the logarithm in Eq.\ \eqref{qdef}.
In a realistic theory,
the running of the QED charge $q$ 
involves all charged fermions,
so the factor $q_0^2$ in Eq.\ \eqref{qdef}
must be replaced with a quantity involving 
a sum over squared charges of all fermions 
participating in the loops.
At sufficiently high energies,
this includes all known fermions in the standard model
and possibly others predicted by the theory
but as yet unobserved.
Any charged scalars in the theory could also play a role.
In many theories,
it may also be necessary to account
for the possible embedding of the charge U(1) group
in a larger unification gauge group.

Since the function $Q(\mu)$ is sensitive to the 
coefficient of the logarithm,
the combination of the above factors 
implies that different physically realistic theories 
can produce a variety of behaviors for $Q(\mu)$.
For the standard-model fermions alone,
the sum of the squared charges is about 0.7,
while the Landau pole for running between
the electroweak scale $m_{ew} \simeq 250$ GeV 
and the Planck scale $M_P \simeq 2 \times 10^{19}$ GeV
occurs at a corresponding factor of about 1.5.
For simplicity and to gain basic insight,
we consider explicitly
the case where this factor is chosen to be 1.

\begin{figure}
\centering
\setlength{\unitlength}{0.0097cm}
\begin{picture}(840,720)(0,0)
\put(20,710){
{\makebox(0,0){$Q^n$}}
}
\put(70,120){
\makebox(780,600)[r]{\incpicwh{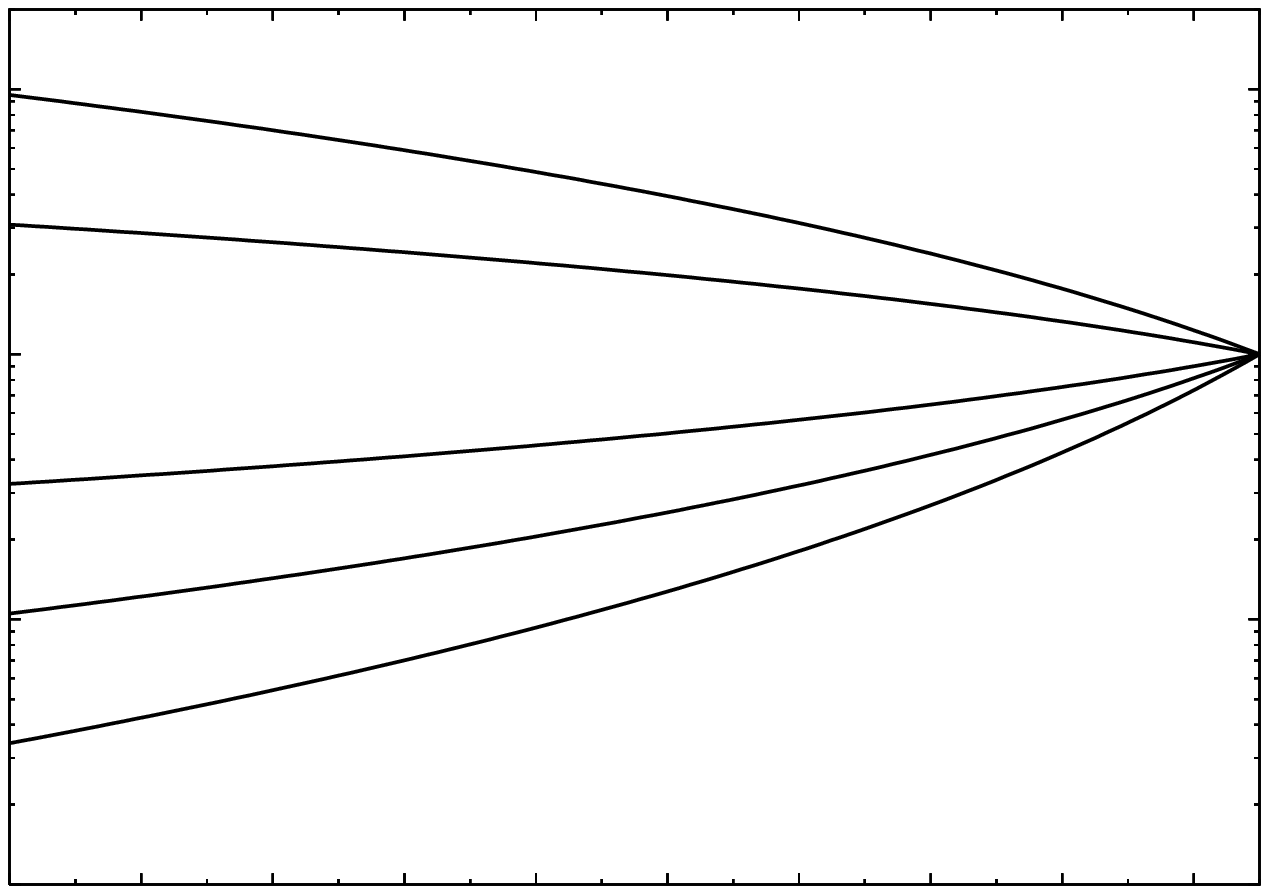}{780\unitlength}{600\unitlength}}
}
\put(60,0){
\makebox(780,60)[c]{$\log (\mu/\text{1~GeV})$}
}
\put(160,120){
\begin{picture}(0,0)
\put(0,90){\makebox(0,0)[l]{$n=-3$}}
\put(0,180){\makebox(0,0)[l]{$n=-2$}}
\put(0,260){\makebox(0,0)[l]{$n=-1$}}
\put(0,470){\makebox(0,0)[l]{$n=1$}}
\put(0,565){\makebox(0,0)[l]{$n=\frac{9}{4}$}}
\end{picture}
}
\put(50,120){
\begin{picture}(0,0)
\put(0,0){\makebox(0,0)[r]{$0.01$}}
\put(0,179){\makebox(0,0)[r]{$0.1$}}
\put(0,362){\makebox(0,0)[r]{$1$}}
\put(0,545){\makebox(0,0)[r]{$10$}}
\end{picture}
}
\put(70,90){
\begin{picture}(0,0)
\put(0,0){\makebox(0,0){$0$}}
\put(82,0){\makebox(0,0){$2$}}
\put(164,0){\makebox(0,0){$4$}}
\put(246,0){\makebox(0,0){$6$}}
\put(328,0){\makebox(0,0){$8$}}
\put(410,0){\makebox(0,0){$10$}}
\put(492,0){\makebox(0,0){$12$}}
\put(574,0){\makebox(0,0){$14$}}
\put(656,0){\makebox(0,0){$16$}}
\put(740,0){\makebox(0,0){$18$}}
\end{picture}
}
\end{picture}
\caption{Variation of the function $Q(\mu)^n$.}
\label{runningpic}
\end{figure}

Figure 6 shows the various powers of the function $Q(\mu)$ 
for this case,
$q_0^2 = 1$,
plotted as a logarithmic function of $\mu$
from $m_{ew}$ to $M_P$.
In a simple scenario for Lorentz and CPT violation,
such as might arise in spontaneous symmetry breaking
\cite{kps},
it is conceivable that 
near the Planck scale all nonzero coefficients for Lorentz violation
are the same order of magnitude in Planck units.
Then,
Fig.\ 6 can be directly interpreted in terms of the divergence 
of renormalization-group trajectories at low energies.
If instead the coefficients for Lorentz violation start at different sizes
at the Planck scale,
then matching the curves in Fig.\ 6 to the trajectories
of the running couplings requires shifts.
In any case,
the figure shows that 
some coefficients for Lorentz violation
increase while others decrease
as the energy scale changes,
with possible relative changes of several orders of magnitude.

The rate of running of the coefficients for Lorentz violation
is relatively small, 
as is to be expected from the logarithmic scale dependence
and from the slow running of the QED coupling $q$.
This running would therefore be insufficient by itself 
to account for the heavy suppression 
of coefficients for Lorentz violation
necessary for compatibility with existing experimental bounds.
If not already present at the Planck scale,
any suppression must be driven by some other factor 
relevant between the electroweak and the Planck scales.
In fact,
it is known that unrenormalizable terms become important
for consistency as the Planck scale is approached
\cite{kle}.
We conjecture that,
in mixing with other coefficients in the renormalization-group equations,
these terms could suffice 
to provide the necessary suppression
at the electroweak scale.
A rapid running 
of coefficients for Lorentz violation with negative mass dimension
is consistent with expectations from the results \rf{runtwo},
which indicate that dimensionless coefficients increase 
towards the Planck scale while the massive ones decrease 
or remain unchanged.
Although it lies outside our present scope,
it would be interesting to investigate further 
this line of reasoning
in the context of explicit models.

In the standard-model extension,
the coefficients for Lorentz violation
may vary with the particle species.
Figure 6 shows that
the running of $Q(\mu)$ with the scale $\mu$
can suffice to induce a significant range of values 
for the various coefficients specific to a given species,
even if all these coefficients are comparable at some large scale.
The figure also suggests that conventional running alone 
cannot separate coefficients for a given species
by more than several orders of magnitude.
In contrast,
for different species of the same nonzero charge,
no separation is induced between 
coefficients for Lorentz violation of a specific type.

Note also that the running of the coefficients
in the full standard-model extension 
must also be controlled by factors
$Q_2(\mu)$ and $Q_3(\mu)$
associated with the SU(2) and SU(3) gauge groups,
respectively.
In some theories,
the SU(3) coupling runs faster than the others,
which suggests a larger spread in the 
quark-sector coefficients for Lorentz violation
and emphasizes the value of a range of tests 
sensitive to different coefficients for hadrons
\cite{kexpt,bexpt,kpcvk,bckp,ccexpt}.
In other theories,
the U(1) coupling runs the fastest,
so the greatest spread may be in the charged-lepton sector,
emphasizing the need for different measurements
with electrons
\cite{eexpt,eexpt2}
and muons
\cite{muexpt}.
Taken together,
all these considerations underline 
the importance of performing experiments 
measuring a wide variety of different coefficients,
both within a given species
and across different sectors of the standard-model extension.

\section{Summary}
\label{conc}

In this paper,
we have shown the one-loop multiplicative renormalizability 
of the general Lorentz- and CPT-violating QED extension \rf{lagdef}.
A generalized Furry theorem has been obtained and used
to prove the absence of one-loop divergences in the cubic
and quartic photon vertices.
The divergent one-loop corrections 
to the photon propagator 
are given in Eq.\ \rf{phtwo},
while those
to the fermion propagator
are in Eqs.\ \rf{qedtwo} and \rf{stfp}.
The associated renormalization factors
are in Eqs.\ \rf{zfall} and \rf{zfacque}.
The divergent one-loop corrections 
to the fermion-photon vertex
are given in Eqs.\ \rf{qedvertex} and \rf{vertwo}.
The usual Ward identities are found to hold
at this order.

We have also initiated an investigation into the
running of the coefficients for Lorentz violation
and the renormalization-group equations.
Following some discussion of the framework,
the one-loop beta functions are obtained 
as Eq.\ \rf{betall}.
The associated partial-differential equations
are solved,
and the running couplings are provided
in Eqs.\ \rf{qdef}, \rf{mqrun}, and \rf{runtwo}.
We show that these equations imply 
that different coefficients for Lorentz violation
typically run differently between
the electroweak and Planck scales
and can lead to a spread of several orders of magnitude
over this range.
Our work emphasizes the value of developing tests
to measure many different coefficients for Lorentz and CPT violation,
both for specific field species
and across all sectors of the standard-model extension.

\section*{Acknowledgments}

This work was supported in part
by the United States Department of Energy
under grant number DE-FG02-91ER40661.
AP wishes to thank John Gracey and D.R.T. Jones for useful discussions.

\appendix

\section*{Feynman rules}

This appendix provides the Feynman rules 
appropriate for our one-loop calculations
using the theory \rf{lagdef}
with the standard gauge-fixing term $-(\prt\cdot A)^2/2\al$.
They are linear in the coefficients for Lorentz and CPT violation.

The fermion propagator is 
\beq
\raisebox{-0.4cm}{\incps{ferm.eps}{-1.5cm}{-.5cm}{1.5cm}{.5cm}}
 &=& i \fr{(\ga_\mu p^\mu  + m )} {p^2 - m^2 },
\eeq
where the momentum $p^\mu$ is understood to 
travel in the direction of the charge arrow shown in the diagram.
The coefficients for Lorentz and CPT violation 
contained in $\Ga_1^\mu$ and $M_1$
lead to insertions in the fermion propagator:
\beq
\raisebox{-0.4cm}{\incps{fermm.eps}{-1.5cm}{-.5cm}{1.5cm}{.5cm}}
&=& - i M_1, \\
\raisebox{-0.4cm}{\incps{fermgam.eps}{-1.5cm}{-.5cm}{1.5cm}{.5cm}} 
&=& i \Ga_1^\mu p_\mu.
\eeq

For a photon of momentum $p^\mu$,
the propagator is 
\beq
\mu \negpad \raisebox{-0.4cm}{\incps{phot.eps}{-1.5cm}{-.5cm}{1.5cm}{.5cm}}
\negpad \nu &=& -
\fr{i}{p^2} ( \et_{\mu\nu} + \fr{p_\mu p_\nu}{p^2} \xi),
\eeq
where $\xi = \al - 1$. 
The coefficients for Lorentz and CPT violation yield 
two types of insertion on this propagator:
\beq
\mu \negpad
\raisebox{-0.4cm}{\incps{photkf.eps}{-1.5cm}{-.5cm}{1.5cm}{.5cm}}
\negpad \nu &=&  -
2 i p^\al p^\be (k_F)_{\al\mu\be\nu}, \\
\mu \negpad
\raisebox{-0.4cm}{\incps{photkaf.eps}{-1.5cm}{-.5cm}{1.5cm}{.5cm}}
\negpad \nu&=& 2
(k_{AF})^\al \ep_{\al\mu\be\nu} p^\be.
\eeq
The momentum $p^\mu$ for the $(k_{AF})^\al$ insertion 
is taken to travel from the $\mu$ index to the $\nu$ index.
Note that the Feynman diagram is symmetric:
the antisymmetry of $\ep_{\al\mu\be\nu}$ 
under $\mu$, $\nu$ interchange
is compensated by the reversal of the momentum direction.

The usual fermion-photon vertex is 
\beq
\raisebox{-.5cm}{\incps{vert.eps}{-1.5cm}{0cm}{1.5cm}{1cm}} &=& -
i q \ga^\mu,
\eeq
where $q$ is the fermion charge 
and $\mu$ is the space-time index on the photon line.
The dimensionless coefficients for Lorentz and CPT violation 
lead to the additional vertex
\beq
\raisebox{-.5cm}{\incps{vertgam.eps}{-1.5cm}{0cm}{1.5cm}{1cm}} 
&=& -i q \Ga_1^\mu.
\eeq

\end{multicols}
\end{document}